\documentclass[jasms]{ametsoc}          % For manuscript

\usepackage{amsfonts}
\usepackage{amsmath}
\usepackage{graphicx}

\lefthead{RYPINA, BERON-VERA, BROWN, KOCAK, OLASCOAGA AND UDOVYDCHENKOV}%
\righthead{SUBMITTED TO THE JOURNAL OF THE ATMOSPHERIC SCIENCES}%
\received{May 5, 2006}

\sectionnumbers
\authoraddr{I.~I. Rypina, RSMAS/AMP, University of Miami, 4600 Rickenbacker
Cswy., Miami, FL 33149 (irypina@rsmas.miami.edu)}       % Corresponding author.

\slugcomment{Submitted to the \emph{Journal of the Atmospheric
Sciences}.}

\begin{document}

\title{On the Lagrangian Dynamics of Atmospheric
Zonal Jets and the Permeability of the Stratospheric Polar Vortex}

\author{I.~I. Rypina, F.~J. Beron-Vera and M.~G. Brown}
\affil{Rosenstiel School of Marine and Atmospheric Science,
University of Miami, Miami, Florida}
\author{H.~Kocak}
\affil{Departments of Computer Science and Mathematics, University
of Miami, Coral Gables, Florida}
\author{M.~J. Olascoaga and I.~A. Udovydchenkov}
\affil{Rosenstiel School of Marine and Atmospheric Science,
University of Miami, Miami, Florida}

\begin{abstract}
The Lagrangian dynamics of zonal jets in the atmosphere are
considered, with particular attention paid to explaining why, under
commonly encountered conditions, zonal jets serve as barriers to
meridional transport.  The velocity field is assumed to be
two-dimensional and incompressible, and composed of a steady zonal
flow with an isolated maximum (a zonal jet) on which two or more
travelling Rossby waves are superimposed. The associated Lagrangian
motion is studied with the aid of KAM (Kolmogorov--Arnold--Moser)
theory, including nontrivial extensions of well-known results. These
extensions include applicability of the theory when the usual
statements of nondegeneracy are violated, and applicability of the
theory to multiply periodic systems, including the absence of Arnold
diffusion in such systems.  These results, together with numerical
simulations based on a model system, provide an explanation of the
mechanism by which zonal jets serve as barriers to meridional
transport of passive tracers under commonly encountered conditions.
Causes for the breakdown of such a barrier are discussed. It is
argued that a barrier of this type accounts for the sharp boundary
of the Antarctic ozone hole at the perimeter of the stratospheric
polar vortex in the austral spring.
\end{abstract}

\begin{article}

\section{Introduction}

It is now generally accepted that the stratospheric polar vortices
in both hemispheres provide effective barriers to meridional
transport of passive tracers.  Although there are differences
between the northern and southern hemispheres and dependence on
height, winds at high latitudes throughout most of the stratosphere
in the winter and early spring are characterized by a nearly zonal
jet; the polar vortex can be defined as the region poleward of the
jet core, and available evidence suggests that the transport barrier
is nearly coincident with the jet core.  The polar vortex in the
northern hemisphere is generally present only in the winter and
early spring.  The stronger southern hemisphere polar vortex often
persists for most of the year, being strongest in the winter and
spring.  Also contributing to the generally stronger southern
hemisphere polar vortex is the observation that Rossby wave
perturbations to the background flow at high latitudes in the middle
and upper stratosphere are generally weaker in the southern
hemisphere than in the northern hemisphere. A more complete
discussion of these topics can be found in \citet{Bowman-93},
\citet{Dahlberg-Bowman-94}, \citet{McIntyre-89} and
\citet{Holton-etal-95}.

Much recent interest in the stratospheric polar vortices derives
from observations of the Antarctic ozone hole and, to a lesser
extent, its northern hemisphere counterpart.  The annual formation
of the Antarctic ozone hole is controlled by chemical processes in
the stratosphere \citep{Webster-etal-93,Lefevre-etal-94}. These
processes will not be discussed here except to note that the
combination of sunlight and cold temperatures that occurs in the
polar region in the early spring triggers ozone depletion. The focus
of the work reported here is explaining the mechanism by which the
stratospheric polar vortex provides a barrier to the meridional
transport of a passive tracer, such as ozone concentration.  Such a
barrier provides an explanation of why, under typical austral early
spring conditions in the middle and upper stratosphere, the ozone
hole does not spread via turbulent diffusion to midlatitudes.

An explanation of the mechanism by which the polar vortex acts as
a barrier to meridional transport has been provided by
\citet{McIntyre-89} and is generally well-accepted.  The argument
assumes that, on a particular isentropic surface, the dynamics are
well approximated by a ``shallow water'' model that conserves
potential vorticity.  The wind field is assumed to be a
superposition of a steady zonal flow with a well-defined maximum
(a zonal jet) and a nonsteady perturbation.  The potential
vorticity distribution associated with the background steady flow
is assumed to be characterized by a strong meridional potential
vorticity gradient at the latitude of the core of the zonal jet.
Because air parcels are constrained to lie on a surface of
constant potential vorticity, the background potential vorticity
gradient will serve as a barrier to meridional transport provided
that the nonsteady perturbation to the vorticity distribution is
not too strong.

In this paper an alternative explanation of the mechanism by which
the polar vortex acts as a barrier to meridional transport is
presented.  Our explanation relies heavily on results relating to
Hamiltonian dynamical systems.  In particular, KAM
(Kolmogorov--Arnold--Moser) theory \citep[see,
e.g.,][]{Arnold-etal-86} plays a central role in our work. Details
will be presented below, but a synopsis of our argument can be given
now. According to the each of many variants of the KAM theorem, if a
steady streamfunction is subjected to certain classes of
time-dependent perturbations, some nonchaotic trajectories (which
lie on tori in a higher dimensional phase space) survive in the
perturbed system. We argue that, under most conditions, the
invariant tori that are most likely to survive in the perturbed
system are those in close proximity to the core of the zonal jet,
and that these provide a barrier to meridional transport.

The connection between KAM tori and the meridional transport barrier
at the perimeter of the stratospheric polar vortex has been
discussed, albeit briefly, in both the meteorological literature
\citep{Pierce-Fairlie-93} and the mathematics literature
\citep{Delshams-delaLlave-00}. Other aspects of dynamical systems
theory have been applied to the stratospheric polar vortices and
described in the meteorological literature \citep[see,
e.g.,][]{Bowman-93,Ngan-Shepherd-99a,Ngan-Shepherd-99b,Koh-Plumb-00,Binson-Legras-02,Koh-Legras-02}.
The relationship between our work and several of these studies will
be discussed below.

There is also a close connection between our work and that of
\citet{Bowman-96}, in spite of the fact that the arguments
presented in that paper are unrelated to dynamical systems.  In
both our work and that of \citet{Bowman-96} the streamfunction is
assumed to consist of a steady background on which a sum of
travelling Rossby waves is superimposed. Bowman's model was based
on an empirical fit to observations.  In addition to this
observational foundation, our model is loosely motivated using
dynamical arguments and chosen, in part, because rigorous
mathematical results are available for streamfunctions of this
general form. Using entirely different arguments than those given
by \citet{Bowman-96}, we provide an explanation for his
observation that for a moderate strength perturbation the
transport barrier in the proximity of the jet core is expected to
break down when one of the Rossby waves included in the
perturbation has a phase speed close to that of the wind speed at
the core of the jet.

The remainder of this paper is organized as follows.  In the next
section, a simple analytic form of the streamfunction is derived.
This consists of a steady background flow -- a zonal jet -- on
which two travelling Rossby waves are superimposed.  In a
reference frame moving at the phase speed of one of the Rossby
waves, the flow consists of a steady background flow on which a
time-periodic perturbation is superimposed.  In section 3, the
Lagrangian motion in such a model is discussed with the aid of two
variants of the KAM theorem. We explain why, under typical
conditions, particle trajectories near the core of the zonal jet
in the perturbed system lie on KAM invariant tori which provide a
barrier to meridional transport.  In section 4, we consider a more
general model of the streamfunction, consisting of a zonal jet on
which three of more travelling Rossby waves are superimposed. The
Lagrangian motion is discussed with the aid of yet another variant
of the KAM theorem. It is argued that the conclusions of section 3
are unchanged for a more general multiperiodic perturbation.  In
section 5 we summarize and discuss our results. Our
KAM-theorem-based explanation of the meridional transport barrier
is contrasted to the potential vorticity barrier explanation, and
suggestions for future work are presented.

\section{A simple, dynamically motivated model of the streamfunction}

Our study focuses on elucidating the mechanism by which the zonal
jet at the edge of the stratospheric polar vortex serves as a
barrier to the meridional transport.  Because of our focus on the
zonal jet, it is natural to make use of a $\beta$-plane
approximation with $\beta = (2\Omega/r_e) \cos \varphi_o$ defined
at the latitude $\varphi_o$ of the core of the zonal jet. Here
$\Omega = 2\pi/(1 {\rm day})$ is the angular frequency of the
earth and $r_e = 6371$ km is the earth's radius.  We shall assume
that $\varphi_o = 60^{\circ}$ so $\beta = 1.14 \times 10^{-11}
{\rm s}^{-1} {\rm m}^{-1}$. Also, our interest is in Lagrangian
motion over time scales of a few months or less. This is
sufficiently short that diabatic processes can be neglected. The
assumption of flow on an isentropic surface, together with
incompressibility, allows the introduction of the streamfunction,
$\psi(x,y,t)$, $u = -\partial \psi/\partial y$, $v =
\partial \psi/\partial x$, with $x$ increasing to the east from an
arbitrarily chosen longitude and $y$ increasing to the north from
$\varphi_o$.  The Lagrangian equations of motion are then
\begin{equation}
\frac{dx}{dt} = -\frac{\partial \psi}{\partial y}, \quad
\frac{dy}{dt} = \frac{\partial \psi}{\partial x}. \label{eq:eq0}
\end{equation}
It is well known that these equations have Hamiltonian form,
$H(p,q,t) \leftrightarrow \psi(x,y,t)$.  This connection is
exploited extensively in sections 3 and 4.

Consistent with our focus on zonal jets we shall assume that
\begin{equation}
\psi(x,y,t) = \psi_0(y) + \psi_1(x,y,t) \label{eq:eq1}
\end{equation}
where $u_0(y) = -\partial \psi_0/\partial y$ has a single extremum
-- a maximum -- at $y=0$. We outline now the steps of a derivation
of a particular choice of $\psi_0(y)$ and $\psi_1(x,y,t)$.  The
same streamfunction has been used previously by
\citet{delCastillo-Morrison-93}, and \citet{Kovalyov-00}. Our
presentation follows that of del-Castillo-Negrete and Morrison;
more details can be found in that work.  The simple analytical
expressions for $\psi_0(y)$ and $\psi_1(x,y,t)$ that are presented
below (see equation \ref{eq:eq11}) are far too simple to mimic the
complexity of realistic stratospheric flows.  But our model of the
streamfunction is dynamically motivated and has approximately the
correct length and time scales.  This model streamfunction is used
to produce numerical simulations to illustrate some important
qualitative features of more realistic flows.  In spite of its
simplicity, our analytic model of the streamfunction includes all
of the essential qualitative features of the stratospheric polar
vortex that are needed to understand why it acts as a meridional
transport barrier.

Consistent with our assumption of 2-d incompressible flow on a
$\beta$-plane, potential vorticity conservation dictates that
\begin{equation}
\frac{\partial \nabla^2 \psi}{\partial t} - \frac{\partial
\psi}{\partial y} \frac{\partial \nabla^2 \psi}{\partial x} +
\frac{\partial \psi}{\partial x} \frac{\partial \nabla^2
\psi}{\partial y} + \beta \frac{\partial \psi}{\partial x} = 0.
\label{eq:eq2}
\end{equation}
Substitution of (\ref{eq:eq1}) into (\ref{eq:eq2}) yields, after
linearization (treating $\psi_1$ as a small perturbation to
$\psi_0$),
\begin{equation}
\frac{\partial}{\partial t} \nabla^2 \psi_1 + u_0(y)
\frac{\partial}{\partial x} \nabla^2 \psi_1 + (\beta - u_0''(y))
\frac{\partial \psi_1}{\partial x} = 0. \label{eq:eq3}
\end{equation}
The assumption that $\psi_1$ has the form of a zonally propagating
wave $\psi_1 = \phi(y)\exp(ik(x-ct))$ (or a superposition of such
waves) yields the Rayleigh--Kuo equation,
\begin{equation}
(u_0(y) - c)(\phi''(y) - k^2 \phi(y)) + (\beta - u_0''(y))\phi(y) =
0. \label{eq:eq4}
\end{equation}
Problems associated with critical layers, where $u_0(y) = c$, and
stability considerations lead to difficulties finding physically
relevant solutions to this equation.  We consider here the Bickley
jet velocity profile
\begin{equation}
u_0(y) = U_0 {\rm sech}^2 \left( \frac{y}{L} \right), \label{eq:eq5}
\end{equation}
where $U_0$ and $L$ are constants.  It was first shown by
\citet{Lipps-62} that for this velocity profile the Rayleigh--Kuo
equation (\ref{eq:eq4}) admits two symmetric neutrally stable
(${\rm Im} \; c = 0$) solutions,
\begin{equation}
\phi_i(y) = A_i U_0 L {\rm sech}^2 \left( \frac{y}{L} \right),
\label{eq:eq6}
\end{equation}
where the $A_i$ $i = 1,2$ are dimensionless amplitudes. It is
straightforward to verify that (\ref{eq:eq5}) and (\ref{eq:eq6})
constitute a solution to (\ref{eq:eq4}) provided
\begin{equation}
U_0 L^2 k^2 = 6c \label{eq:eq7}
\end{equation}
and
\begin{equation}
6c^2 - 4 U_0 c + \beta U_0 L^2 = 0. \label{eq:eq8}
\end{equation}
The condition for the existence of two neutrally stable waves is
\begin{equation}
\beta L^2/U_0 < 2/3. \label{eq:eq9}
\end{equation}
When this inequality is satisfied, (\ref{eq:eq8}) has two real
roots; the corresponding wavenumbers are given by (\ref{eq:eq7}).
It should be noted that for these solutions to (\ref{eq:eq4}),
$(\beta - u_0''(y))/(u_0(y) - c)$ is bounded at the critical
layers; this is a necessary condition for the existence of
neutrally stable solutions \citep{Kuo-49}.

The environment is defined by the parameters $\beta$, $U_0$ and $L$,
which, via (\ref{eq:eq7}) and (\ref{eq:eq8}), fix $c_1$, $k_1$,
$c_2$, and $k_2$ provided (\ref{eq:eq9}) is satisfied. But, because
of the periodic boundary conditions in $x$, only a discrete set of
$k$ are allowed.  At $\varphi_0 = 60^{\circ}$ (this choice fixes
$\beta$ as noted above) these are
\begin{equation}
k_n = \frac{2n}{r_e}, \quad n = 1,2, \ldots .  \label{eq:eq10}
\end{equation}
For most choices of $U_0$ and $L$, (\ref{eq:eq10}) conflicts with
(\ref{eq:eq7}) and (\ref{eq:eq8}).  This issue was discussed by
\citet{delCastillo-Morrison-93} who argued that initial
disturbances for which (\ref{eq:eq7}) and (\ref{eq:eq8}) are
inconsistent with (\ref{eq:eq10}) should relax, via a
barotropic-instability-induced decrease in $U_0$ and increase in
$L$, to a state for which (\ref{eq:eq7}), (\ref{eq:eq8}) and
(\ref{eq:eq10}) are self-consistent.  We avoid this issue by
choosing $U_0$ and $L$ that correspond to such a self-consistent
state.  Specifically, we have chosen $U_0 = 62.66$ m/s, $L = 1770$
km, corresponding to zonal wavenumbers $n = 2$ and $n = 3$.  These
waves have eastward propagating phase speeds, $c_2/U_0 = 0.205$
and $c_3/U_0 = 0.461$. The streamfunction is then
\begin{eqnarray}
\psi(x,y,t) = -U_0 L \tanh \left(\frac{y}{L}\right)\!&+&\!A_3U_0 L
\, {\rm sech}^2 \left(\frac{y}{L}\right) \cos(k_3(x-c_3t)) \nonumber \\
&+&\!A_2 U_0 L \, {\rm sech}^2 \left(\frac{y}{L}\right)
\cos(k_2(x-c_2t)). \label{eq:eq11}
\end{eqnarray}
An important observation is that the time dependence associated with
one of the two Rossby waves in (\ref{eq:eq11}) can be eliminated by
viewing the flow in a reference frame moving at the phase speed of
that wave.  The choice of which wave to absorb into the background
flow is arbitrary.  In the reference frame moving at speed $c_3$ the
streamfunction is
\begin{eqnarray}
\psi(x,y,t) = c_3 y - U_0L\tanh\left(\frac{y}{L}\right)\!&+&\!A_3
U_0 L \, {\rm sech}^2 \left(\frac{y}{L}\right) \cos(k_3x) \nonumber \\
&+&\!A_2 U_0 L \, {\rm sech}^2 \left(\frac{y}{L}\right)
\cos(k_2x-\sigma_2t) \label{eq:eq12}
\end{eqnarray}
where $\sigma_2 = c_2 k_2 - c_3 k_2 = k_2 (c_2 - c_3)$. Note that
$\sigma_2$ is negative because in the reference frame moving at the
faster $n = 3$ wave, the $n = 2$ wave has westward propagating
phases.

\section{A steady background flow subject to a periodic perturbation}

In this section we consider streamfunctions of the general form
\begin{equation}
\psi(x,y,t) = \psi_0(x,y) + \psi_1(x,y,\sigma t) \label{eq:eq13}
\end{equation}
where $\psi_1$ is a periodic function of $t$ with period
$2\pi/\sigma$.  Equation (\ref{eq:eq12}) is a special case of
equation (\ref{eq:eq13}).   All of the concepts discussed in this
section apply to the more general class (\ref{eq:eq13}).  The
particular form (\ref{eq:eq12}) is used for numerical simulations
to illustrate the relevant important concepts. In the section that
follows we consider a slightly larger and more geophysically
relevant class of streamfunctions corresponding to a multiperiodic
perturbation $\psi_1$.  It will be seen that almost all of the
results presented in this section generalize in a straightforward
fashion to multiperiodically perturbed systems.

We begin with a discussion of the importance of the background
steady contribution to the streamfunction $\psi_0(x,y)$ in
(\ref{eq:eq13}).  If, as we have assumed, in the rest frame the
streamfunction has the form of a zonal flow on which a sum of
zonally propagating Rossby waves are superimposed (as in
(\ref{eq:eq11})), the problem of lack of uniqueness of
$\psi_0(x,y)$ arises immediately; the choice of which travelling
wave to absorb into the background is arbitrary.  Because the
purely zonal rest frame contribution to $\psi_0(x,y)$ ($-U_0 L
\tanh (y/L)$ in (\ref{eq:eq12})) is always present and is
generally larger than whichever travelling wave contribution is
absorbed into $\psi_0$, we discuss first the special case $\psi =
\psi_0(y)$.

The Hamiltonian form of the Lagrangian equations of motion
(\ref{eq:eq0}) was noted earlier.  The special case $\psi =
\psi_0(y)$ corresponds, trivially, to the so-called action-angle
representation of the motion in which $(p,q) \rightarrow
(I,\theta)$, $H(p,q) \rightarrow \bar{H}(I)$.  The equations of
motion in terms of action-angle variables $(I,\theta)$ are $dI/dt
= - \partial \bar{H}/\partial \theta = 0$, $d\theta/dt =
\partial \bar{H}/\partial I \equiv \omega(I)$; these equations
can be trivially integrated.  Note that $I$ and $\theta$ are
defined in such a way that the motion is $2\pi$-periodic in
$\theta$ with angular frequency $\omega(I)$.  When $\psi =
\psi_0(y)$, we may take $I = -yR$, $\theta = x/R$ and $\bar{H} =
\psi_0$, where $R = r_e \cos \varphi_o$.  With these simple
substitutions the original Lagrangian equations of motion
(\ref{eq:eq0}) have action-angle form.  For systems of this type
$\omega(I)$ is simply a relabelling of $u(y)$ and $T(I) =
2\pi/\omega(I)$ is the time required for a trajectory to circle
the earth.  Action-angle variables and, in particular, the
quantity $d\omega(I)/dI$ play a crucial role in much of the
discussion that follows.

The choice $\psi_0(y) = - U_0 L \tanh(y/L)$ corresponds to
$\bar{H}(I) = U_0 L \tanh(I/(RL))$, $\omega(I) = \partial
\bar{H}/\partial I = (U_0/R) \, {\rm sech}^2(I/(RL))$.  Figure 1a
shows the corresponding streamlines in the $(x,y)$-plane (the
trivial $x$-dependence is included for comparison to Fig. 3), and
plots of $u_0(y) = - \partial \psi_0/\partial y = U_0 \, {\rm
sech}^2 (y/L)$, $\omega(I)$ and $\omega'(I)$. Note that at the jet
core, $T(I)$ has a local minimum, $\omega(I)$ has a local maximum
and $\omega'(I) = 0$.

\begin{figure}[t]
\centering
\includegraphics[width=16cm]{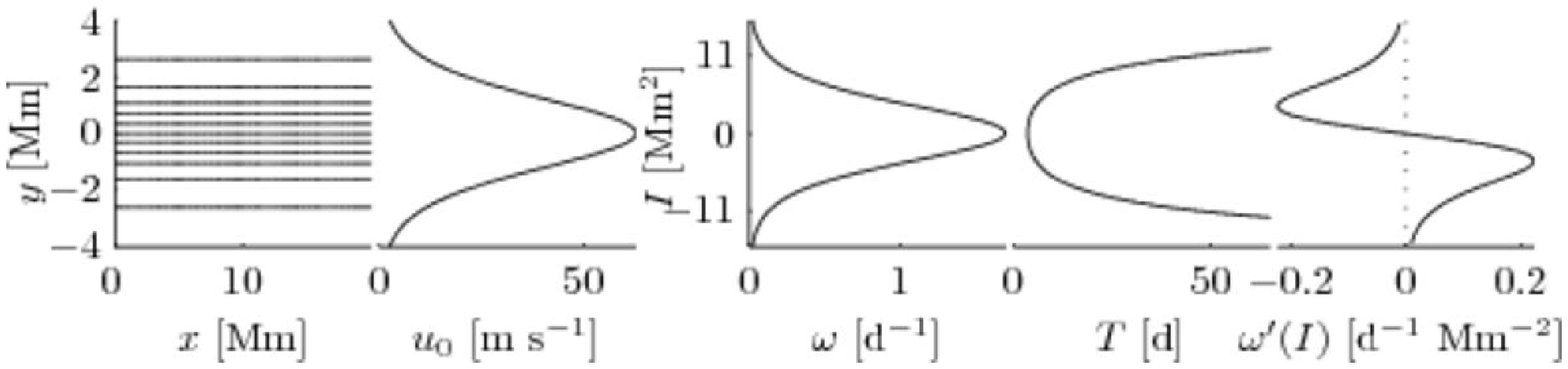}
\caption{For the streamfunction $\psi_0(y) = -U_0 L \tanh(y/L)$: (a)
selected level surfaces of $\psi_0$; (b) $u_0(y)$; (c) $\omega(I)$;
(d) $T(I)$; (e) $\omega'(I)$.  In this and subsequent figures, d
denotes days and distance is measured in Mm; 1 Mm = 1000 km.}
\end{figure}

Now consider a superposition of the background zonal jet $\psi_0(y)$
and one of the two Rossby wave perturbations included in
(\ref{eq:eq11}).  In the reference frame moving at the phase speed
of the Rossby wave the flow is steady.  The corresponding
streamfunction is
\begin{equation}
\psi(x,y) = c y - U_0 L \tanh \left(\frac{y}{L}\right) +  A U_0 L \,
{\rm sech}^2 \left(\frac{y}{L}\right) \cos(kx) \label{eq:eq14}
\end{equation}
where the phase speed $c$, wavenumber $k$ and the dimensionless
wave amplitude $A$ are written without subscripts.
Nondimensionalization ($\psi \rightarrow \psi/(U_0 L), x
\rightarrow kx, y \rightarrow y/L$) reveals that there are two
irreducible dimensionless parameters, $A$ and $c/U_0$.  A
bifurcation diagram in $(A,c/U_0)$ for this system is shown in
Fig. 2.  This figure shows that there are three regions,
corresponding to topologically distinct streamfunction structures,
and two critical curves that separate these regions. Level
surfaces of $\psi$ in each of the three regions and on the two
critical lines are shown in the figure. Holding $A \neq 0$
constant while $c/U_0$ is increased reveals all possible
streamfunction topologies. For small $c/U_0$ the streamfunction is
characterized by hyperbolic heteroclinic chains both above and
below a spatially periodic eastward jet near $y=0$. As $c/U_0$ is
increased, a critical value is encountered, at which the two
hyperbolic heteroclinic chains merge and the eastward jet
disappears.  A further increase of $c/U_0$ leads to the formation
of homoclinic hyperbolic chains above and below a westward jet. As
a second critical value of $c/U_0$ is passed, the hyperbolic
homoclinic chains are destroyed via saddle-node annihilation. For
large $c/U_0$ the flow is everywhere westward without stagnation
points. Similar behavior was noted previously by
\citet{delCastillo-Morrison-93} using essentially the same model.
The importance of Fig. 2 is that it shows that depending on the
choice of $A$ and $c/U_0$, the zonal jet may be strong, weak or
absent entirely. If $A$ and $c/U_0$ correspond to a pair which
lies on the critical line at which the two hyperbolic heteroclinic
chains merge, the eastward jet disappears and the chain of
unstable and stable manifolds near $y = 0$ is unstable to an
arbitrarily small time-dependent perturbation. For the
stratospheric polar vortex the relevant (usually) domain of the
$(A,c/U_0)$ parameter space is small values of both parameters
(see, e.g., Bowman, 1996). It should be emphasized, however, that
when more than one Rossby wave is superimposed on the background
zonal jet, as in (\ref{eq:eq12}), the choice of which Rossby wave
to absorb into the background is arbitrary. Under such conditions
the claim that the background flow topology corresponds to the
small $c/U_0$ region in Fig. 2 is justified only if this is true
for $(A_i,c_i/U_0)$ for all of the waves present. Although the
preceding discussion was motivated by a particular model
streamfunction, equation (\ref{eq:eq14}), the qualitative features
that we have described are expected to be broadly applicable to
Rossby wave perturbations to zonal jets.

\begin{figure}[t]
\centering
\includegraphics[width=14cm]{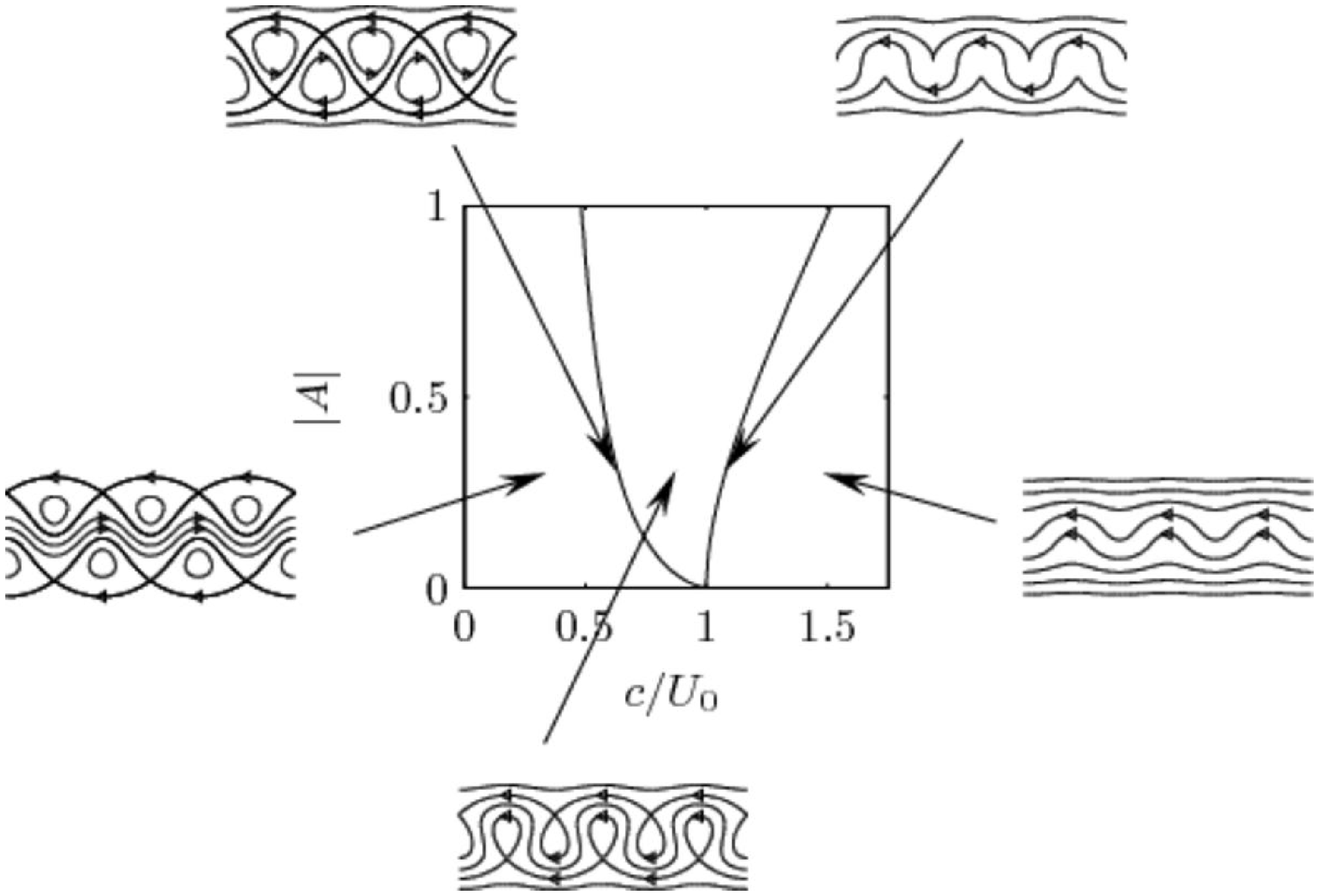}
\caption{Bifurcation diagram in the $(A,c/U_0)$ parameter space
corresponding to the streamfunction $\psi(x,y) = cy - U_0 L
\tanh(y/L) + A U_0 L {\rm sech}^2(y/L)\cos(kx)$.  There are three
topologically distinct regions and two critical curves separating
these regions. Selected level surfaces of $\psi(x,y)$ in each of the
three regions and on the two critical curves are shown. In the level
surface plots zonal wavenumber three is assumed so $k = k_3$, and $0
\leq k_3x \leq 6\pi$.}
\end{figure}

For any steady streamfunction $\psi = \psi(x,y)$ the Lagrangian
equations of motion (\ref{eq:eq0}) can be transformed to
action-angle form. The equations of motion in action-angle form
are identical to the equations described above, but the
transformation from $\psi(x,y)$ to $\bar{H}(I)$ is more
complicated than the trivial relabelling of coordinates described
above.  More generally, $I(\bar{H}) = (2\pi)^{-1}\oint
x(y,\bar{H}) dy = -(2\pi)^{-1}\oint y(x,\bar{H}) dx$ where
$\bar{H} = \psi$ and the integral is around a closed loop in
$(x,y)$, and $\theta = \partial G/\partial I$ with $G(y,I) =
\int_0^y x(y',\bar{H}) dy'$.  (The equivalence of the two forms of
$I(\bar{H})$ given above follows from integration by parts.  Note
also that on a given level surface of $\psi$, $x(y)$ or $y(x)$ may
be multi-valued, dictating that some care be exercised when using
these equations, and that there is flexibility in choosing the
lower limit in the integral defining $G$.)  It is often necessary
to define action-angle variables in different regions of $(x,y)$
in a piecewise fashion. We emphasize, however, that once this is
done the form of the equations of motion in action-angle variables
is that given above. It is important to keep in mind that $I$ is
simply a label for a particular trajectory or, equivalently, for a
particular level surface of $\psi(x,y)$.

For $A = 0.3$, $c/U_0 = 0.461$, plots of $|{\bf u}| = (u^2 +
v^2)^{1/2}$, $\omega(I)$, $T(I)$ and $\omega'(I)$ are shown in
Fig. 3 for trajectories in the vicinity of the jet core only. Note
that, in qualitative agreement with Fig. 1, Fig. 3 shows that in
the vicinity of the jet core, $T(I)$ has a local minimum,
$\omega(I)$ has a local maximum and $\omega'(I) = 0$. These
features play an important role in the considerations that follow.

\begin{figure}[t]
\centering
\includegraphics[width=16cm]{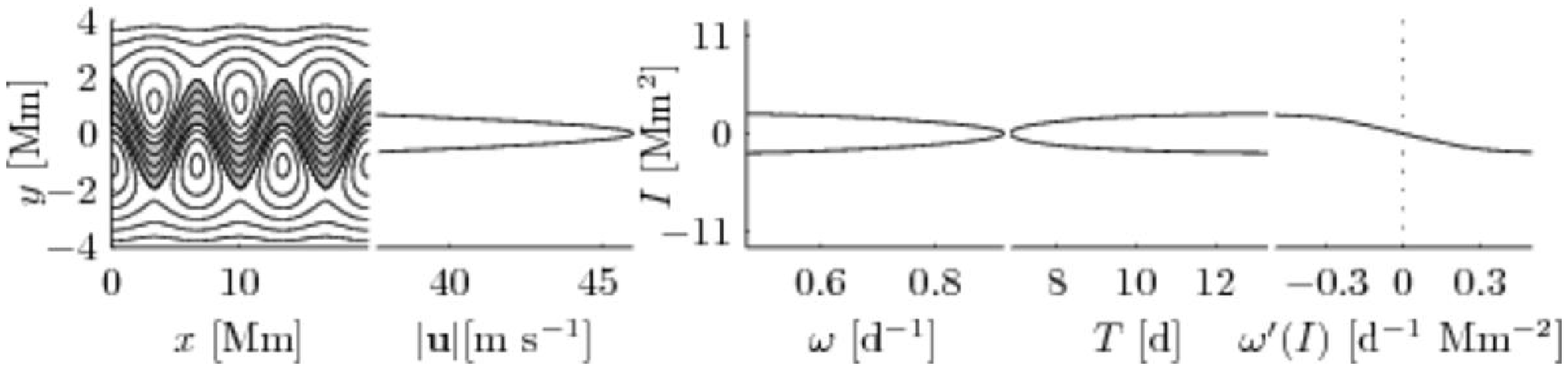}
\caption{For the streamfunction $\psi(x,y) = c_3y - U_0 L \tanh(y/L)
+ A_3 U_0 L {\rm sech}^2(y/L)\cos(k_3x)$ with $A_3 = 0.3$, $c_3/U_0
= 0.461$: (a) selected level surfaces of $\psi(x,y)$; (b) $|{\bf
u}|(y)$ at $k_3x = \pi/2$; (c) $\omega(I)$; d) $T(I)$; and (e)
$\omega'(I)$. In (b), (c), (d) and (e) only values of $y$ and $I$
corresponding to the shaded region near the jet core in (a) are
shown.}
\end{figure}

We turn our attention now to periodically perturbed systems of the
form (\ref{eq:eq13}), of which (\ref{eq:eq12}) is a special case.
With $x$ and $y$ bounded, trajectories lie in a 3-dimensional
bounded phase space $(x,y,t \bmod 2\pi/\sigma)$. The usual way to
view trajectories in such a system is to construct a Poincare
section, which is a slice of the 3-d space corresponding to $t
\bmod 2\pi/\sigma = {\rm constant}$. Three examples, corresponding
to the system described by (\ref{eq:eq12}) with three choices of
the perturbation strength $A_2$, are shown in Fig. 4.  On these
plots regular (nonchaotic) trajectories appear as discretely
sampled smooth curves, while chaotic trajectories appear as sets
of discrete samples that fill areas.  In the $A_2 = 0$ limit all
trajectories are nonchaotic; each curve seen in Fig. 4a can be
thought of as a 2-d slice of a torus in $(x,y,t \bmod
2\pi/\sigma)$. For small perturbation strength $A_2$ some of the
unperturbed tori are seen to survive, while other tori break up
forming chains of island-like structures that are surrounded by
chaotic seas.  In general, as the perturbation strength increases
more tori are destroyed and the motion becomes increasingly
chaotic.

\begin{figure}[t]
\centering
\includegraphics[width=14cm,angle=-90]{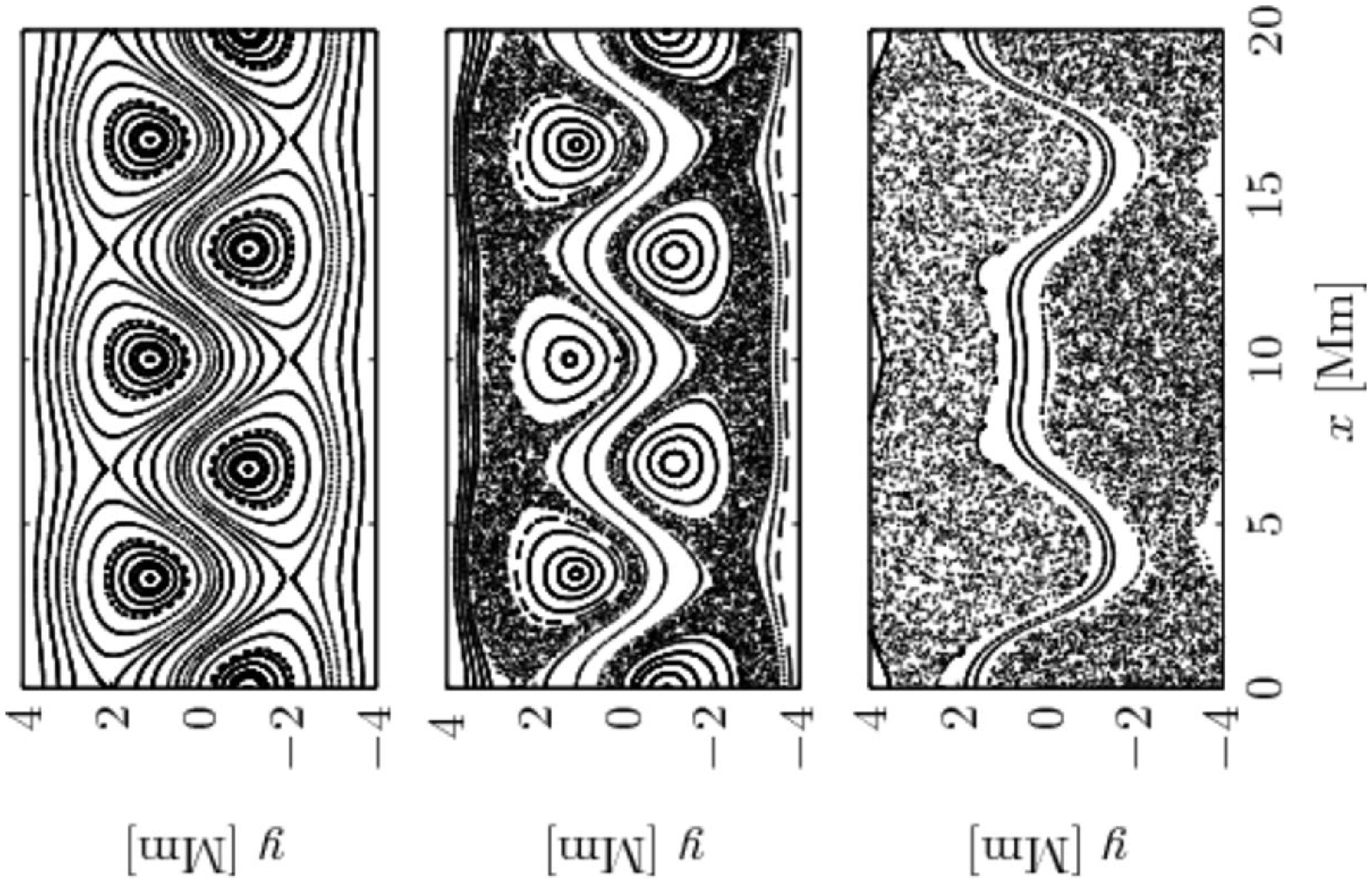}
\caption{Poincare sections corresponding to the system described
by equations (\ref{eq:eq0}) and (\ref{eq:eq12}) with $A_3 = 0.3$
for three values of $A_2$: $0$ (upper plot), $0.1$ (middle plot),
and $0.7$ (lower plot).  Note the robustness of the tori in the
vicinity of the jet core.}
\end{figure}

Before proceeding, it is instructive to make some seemingly
trivial comments about the geometry of systems of the form
(\ref{eq:eq13}). The tori of the unperturbed system can be thought
of as either 1-d surfaces in $(x,y)$ or 2-d surfaces in the 3-d
space $(x,y,t \bmod 2\pi/\sigma)$.  Because the unperturbed
streamfunction does not depend on $\sigma$ the latter view seems
like an unnecessary complication, but it turns out to be very
useful.  The regular trajectories in the perturbed systems shown
in Fig. 4 lie on tori (of the second type) of the unperturbed
system that survive under perturbation.  Because each surviving
torus is a 2-d surface in the 3-d space $(x,y,t \bmod
2\pi/\sigma)$, each such torus divides the 3-d space into disjoint
`inside' and `outside' regions.  A consequence of this is that in
the $(x,y)$-plane each surviving torus represents an impenetrable
barrier to transport.

The survival of some of the tori of the unperturbed system under
small perturbation, as illustrated in Fig. 4, is predicted by the
KAM theorem \citep[see, e.g.,][]{Arnold-etal-86,Poschel-01}.  Before
giving a more precise statement of the theorem, it is instructive to
note that the mechanism that leads to the destruction of the tori of
the unperturbed system is excitation of resonances by the
time-periodic perturbation. Resonances are excited when the ratio of
the frequency of the perturbation $\sigma$ to the frequency of the
motion on the unperturbed torus $\omega(I)$ is the ratio of
integers. Generically, a continuum of $\omega(I)$ values are
present.  Under such conditions a fixed $\sigma$ excites infinitely
many resonances. In practice, however, the low-order resonances
(e.g., $2:1$) are the most important.

With this as a heuristic background, a form of the KAM theorem
suitable for systems of the form (\ref{eq:eq13}) can now be stated.
According to the theorem, tori in the vicinity of those tori in the
unperturbed system for which $\omega(I)/\sigma$ is sufficiently
irrational survive in the perturbed system provided the strength of
the perturbation is sufficiently weak and a nondegeneracy condition,
$\omega'(I) \neq 0$, is satisfied.  The condition that
$\omega(I)/\sigma$ is sufficiently irrational is expressed
quantitatively by a Diophantine condition; this will not be
discussed further as this condition is not central to our arguments.

On the other hand, the nondegeneracy condition plays a critical role
in our arguments. In the simplest form of the theorem this condition
is $\omega'(I) \neq 0$ (or in higher dimensions ${\rm det}(\partial
\omega_i/\partial I_j) \neq 0$).  This condition guarantees the
invertibility of $\omega(I)$, whose importance stems in part from
the fact that the theorem guarantees that the torus corresponding to
that value of $I$ for which $\omega(I)/\sigma$ is sufficiently
irrational survives in the perturbed system.  This was the form of
the nondegeneracy condition in Kolmogorov's (1954)
\nocite{Kolmogorov-54} original statement of the theorem. Already in
his original proof of the theorem, \citet{Arnold-63} noted (in a
footnote) that an alternate form of the nondegeneracy condition, the
isoenergetic condition, could be used instead. Subsequently,
\citet{Bruno-92} and \citet{Russmann-89} announced forms of the
theorem that employ less restrictive nondegeneracy conditions.

The Russmann nondegeneracy condition is of particular interest in
the present study.  The condition is most naturally stated in
words: for an autonomous system with $N+1$ degrees of freedom the
image of the frequency map $I \rightarrow \omega(I)$ may not lie
on any hyperplane of dimension $N$ that passes through the origin.
To our knowledge, all published formulations of the KAM theorem to
date that make use of the Russmann nondegeneracy condition apply
to autonomous systems. To apply such a result to (\ref{eq:eq13})
this system must first be written as an equivalent autonomous two
degree-of-freedom system with a bounded phase space. The required
transformation is a special case of the transformation described
at the end of the following section. After performing this
transformation, the Russmann condition reduces to a statement that
in the 2-d $(\sigma,\omega)$-space, the locus of points
$(\sigma,\omega(I))$ must not fall on a line that passes through
the origin. This condition is violated only when $\omega(I) =
\omega_o$, a constant.  For our purposes, the significance of the
Russmann nondegeneracy condition is that, for systems of the form
(\ref{eq:eq13}), it is satisfied in a domain that includes an
isolated zero of $\omega'(I)$.   The price paid for making use of
the relatively weak Russmann nondegeneracy condition is that a
slightly less strong form of the theorem is proved.  As noted
above, the Kolmogorov condition $\omega'(I) \neq 0$ can be used to
prove that the torus corresponding to a particular value of $I$ in
the unperturbed system survives in the perturbed system. In
contrast, when the Russmann form of the theorem is applied in the
vicinity of a torus for which $I = I_0$ where $\omega'(I_0) = 0$
(an isolated zero) the theorem guarantees only that some tori
identified by $I$-values near $I_0$ will be present in the
perturbed system. Thus, when the Russmann form of the theorem is
applied it is not appropriate to refer to surviving tori
\citep[see, e.g.,][for a more complete discussion of this
issue]{Sevryuk-95,Sevryuk-06}. For our purposes this distinction
is unimportant in that it makes no difference whether the $I =
I_0$ torus survives under perturbation; it is important to know
only that some nearby tori are present in the perturbed system as
any such torus provides a barrier to transport.

Because of its generality, we have chosen to emphasize the
importance of the Russmann nondegeneracy condition in our
discussion of the stability of tori in the vicinity of that for
which $\omega'(I) = 0$. It is worth noting, however, that other
arguments have been used to establish essentially the same result
for area-preserving mappings
\citep{Delshams-delaLlave-00,Simo-98}.

Consider again the Poincare sections shown in Fig. 4.  This figure
shows that not only do the tori corresponding to trajectories near
the jet core (where $\omega'(I) = 0$) persist in the perturbed
system, but these tori appear to be the most resistant to
breaking. Numerical simulations based on model systems reveal
that, in general, tori near that for which $\omega'(I) = 0$ tend
to be the most resistant to breaking. Interestingly, for the model
parameters used to construct Fig. 4, $\sigma/\omega$ at the jet
core is approximately $0.95$.  The significance of this ratio is
its closeness to unity.  On two nearby tori, one on each side of
the jet core, the strongest possible $(1:1)$ resonance is excited.
In spite of this, tori in the vicinity of the jet core are seen to
be preserved for moderate perturbation strengths.  The reason for
the surprising stability of tori near that for which $\omega'(I) =
0$ will be described below. We emphasize, however, that the
stability of tori near that for which $\omega'(I) = 0$ is not
absolute. If, for instance, $\sigma$ happens to be identical to
$\omega(I)$ at the jet core where $\omega'(I) = 0$, thereby
exciting a $1:1$ resonance on the jet core, tori near the jet core
are not among the last to break up as the perturbation strength
increases.

The fact that tori for which $\omega'(I)=0$ are strongly resistant
to breaking has been noted in the mathematical literature;
\citet{Gaidashev-Koch-04} refer to the ``remarkable stability'' of
such tori. Systems which satisfy this condition are generally
described as ``shearless'' or ``nontwist'' in the mathematical
literature, and have been extensively studied in recent years
\citep[see,
e.g.,][]{delCastillo-Morrison-93,Morozov-02,Dullin-Meiss-03}.

We turn our attention now to resonance widths as a means to
explain the remarkable stability of tori satisfying the nontwist
condition. Resonance widths are important because when neighboring
resonances overlap, the intervening tori generally break up; the
widely used Chirikov definition of chaos is based on overlapping
resonances \citep[see,
e.g.,][]{Chirikov-79,Chirikov-Zaslavsky-72,Lichtenberg-Lieberman-83}.
Recall that resonances are excited on tori for which
$\omega(I)/\sigma$ is rational. Resonance widths are controlled by
the degree of rationality of $\omega/\sigma$, the perturbation
strength and a simple geometric factor, which we now consider.  A
simple analysis \citep[see, e.g.,][]{Abdullaev-93b} reveals that
resonance widths scale like $\Delta I \sim |\omega'(I)|^{-1/2}$,
or $\Delta \omega \sim |\omega'(I)|^{1/2}$. Because resonances are
excited at discrete values of $\omega$, it is the width $\Delta
\omega$, rather than $\Delta I$, that is important in determining
whether neighboring resonances overlap.  Because $\Delta \omega
\sim |\omega'(I)|^{1/2}$ small values of $\omega'(I)$ are
generally associated with small resonance widths, and generally
more surviving KAM tori.  (The resonance width estimates just
quoted follow from a simple perturbation analysis.  When
$\omega'(I) = 0$ at the resonance, the width of the resonance
$\Delta \omega$ depends on $\omega''(I)$ at the resonance.  The
exact form of this expression is not essential to our argument.
What is important is the observation that $\Delta \omega$ is small
when $\omega'$ is small.)

In the vicinity of the jet core a narrow band of $\omega$-values
will be present.  Resonances will be excited in this band, but
only for very special values of $\sigma$ will these be low-order
resonances. The associated widths of these resonances are small
owing to the smallness of $|\omega'(I)|^{1/2}$ in this region.  As
a result, mostly nonchaotic motion is preserved near the jet core,
not because resonances are not excited, but because the
corresponding resonance widths are usually so small that
neighboring resonances don't overlap. Excitation of a low-order
resonance very close to the jet core can overcome the smallness of
$|\omega'(I)|^{1/2}$ and change this picture, so the stability of
tori near the jet core is not absolute.

In this section we have considered a steady zonal jet on which two
travelling Rossby waves are superimposed.  Either of the two
Rossby waves can be absorbed into a modified steady background
flow.  We have shown, using well-known results relating to KAM
theory, that, provided certain conditions are met, a typically
narrow band of nonchaotic trajectories in the vicinity of the jet
core, each lying on a KAM invariant torus, persists in the two
wave system and provides a barrier to meridional transport.  The
barrier is linked to the remarkable stability of KAM tori for
which $\omega'(I)$ has an isolated zero.  The conditions that need
to be met for such a barrier to be present are: (1) the rest frame
phase speeds of both Rossby waves should not be comparable to the
wind speed at the jet core; (2) the Rossby wave amplitudes must
not be too large; and (3) low order resonances in the immediate
vicinity of the jet core in the moving frame must not be excited.

\section{A steady background flow subject to a multiperiodic perturbation}

In this section we consider streamfunctions of the form
\begin{equation}
\psi(x,y,t) = \psi_0(x,y) + \psi_1(x,y,\sigma_1 t, \ldots, \sigma_N
t) \label{eq:eq16}
\end{equation}
where $\psi_1$ is a multiperiodic function with constituent
periods $2\pi/\sigma_i, i = 1,2,\ldots,N$. It should be noted that
a steady zonal flow on which a sum of $N+1$ zonally propagating
Rossby waves is superimposed has the form (\ref{eq:eq16}) when
viewed in the reference frame moving at the phase speed of one of
the Rossby waves. The $N = 1$ problem treated in the previous
section is seen to be a special case of the problem treated here.
In this section we show that most of the results discussed in the
previous section carry over to the larger and more realistic class
of problems considered here with only minor modification.

An important observation relating to systems of the form
(\ref{eq:eq16}) is that one need only consider frequencies that
are incommensurable, i.e., have the property that the ratio of all
pairs of frequencies is irrational. Consider, for example, a
multiperiodic function with periods 4 and 6 days. This function is
a simple periodic function with period 12 days. In general, a
reduction of the number of frequencies can be achieved whenever
two more of the frequencies are commensurable. Thus, without loss
of generality, it may be assumed that $\sigma_1, \sigma_2, \ldots,
\sigma_N$ are incommensurable, i.e., that $\psi_1$ is a
quasiperiodic with $N$ incommensurable frequencies.

Systems of the form (\ref{eq:eq16}) have been intensively studied
in recent years.  A proof of the KAM theorem for such systems has
been provided by \citet{Jorba-Simo-96}. Several points relating to
the Jorba--Simo work are noteworthy. First, the theorem is
formulated as a nonautonomous perturbation to an autonomous one
degree of freedom system, so the unperturbed Hamiltonian, which
must satisfy a nondegeneracy condition, is the system defined by
$\psi_0(x,y)$ (after transforming to the action-angle
representation).  Second, the nondegeneracy condition that the
unperturbed Hamiltonian is assumed to satisfy is the Kolmogorov
condition $\omega'(I) \neq 0$. Third, Diophantine conditions must
be satisfied by both $\sigma_i/\omega$ and $\sigma_i/\sigma_j$ $(i
\neq j)$.  The second point is of particular importance in the
present study.  Loosely speaking, the Jorba--Simo work shows that
the principal difference between the periodic perturbation problem
and the quasiperiodic perturbation problem is that in the former
problem the surviving KAM tori undergo periodic oscillations in
$(x,y)$, while in the latter problem the surviving KAM tori
undergo quasiperiodic oscillations in $(x,y)$.  Jorba and Simo
refer to the latter motion as a ``quasiperiodic dance.'' For our
purposes, this distinction is unimportant; in both cases the
surviving KAM tori provide a barrier in $(x,y)$ to transport, as
we shall describe in more detail later in this section.

All of the mathematical difficulties associated with a
quasiperiodic perturbation $\psi_1$ are present even for $N = 2$.
Because, among all $N \geq 2$, the $N = 2$ case is the most
convenient choice for numerical purposes, it is natural to focus
on that choice.  With this in mind we have chosen, for numerical
purposes, to use the streamfunction
\begin{multline}
\psi(x,y,t) = c_3 y - U_0 L \tanh \left(\frac{y}{L}\right) + A_3
U_0 L {\rm sech}^2 \left(\frac{y}{L}\right) \cos(k_3x) \\
+ A_2 U_0 L{\rm sech}^2 \left(\frac{y}{L}\right)
\cos(k_2x-\sigma_2t) + A_1 U_0 L {\rm sech}^2
\left(\frac{y}{L}\right) \cos(k_1x-\sigma_1t). \label{eq:eq17}
\end{multline}
In the $A_1 = 0$ limit this is identical to the streamfunction
described by equation (\ref{eq:eq12}).  Note that physically
equation (\ref{eq:eq17}) represents a zonal flow corresponding to
$\psi(y) = - U_0 L \tanh (y/L)$ on which three travelling
Rossby-like waves are superimposed in a reference frame moving
with speed $c_3$, the phase speed of the zonal wavenumber three
wave. For convenience, we have assumed that the new perturbation
term corresponds to zonal wavenumber one, $k_1 = 2\pi/(2\pi r_e
\cos 60^{\circ}) = 2/r_e$, and has the same ${\rm sech}^2(y/L)$
meridional structure as the $k_2$ and $k_3$ modes.  However,
unlike the $k_2$ and $k_3$ modes, which had some dynamical
justification, the $k_1$ mode is simply an ad-hoc additive
perturbation which is included to illustrate some properties of
quasiperiodic systems. With this in mind we have chosen
$\sigma_1/\sigma_2$ to be the golden mean $(\sqrt{5}-1)/2$ (whose
continued fraction representation identifies it as the most
irrational real number).

Numerical simulations based on the system defined by
(\ref{eq:eq17}) are shown in Figs. 5 and 6.  Figure 5 shows the
time evolution of two sets of air parcels at times ranging from $t
= 0$ to $t =81$ days. The initial conditions are chosen to fall on
two zonal lines $y = {\rm constant}$ on opposite sides of the
zonal jet.  It is seen that after 81 days each side of the jet is
well stirred, as indicated by what appears to be random
distributions of dots on each side of the jet, but there is no
transport across a wavy boundary near the core of the jet.  The
cause of this behavior is a thin band of KAM invariant tori near
the jet core that survive in the perturbed system and form a
meridional transport barrier. This thin band of KAM invariant tori
that separate the polar from the midlatitude region in our
idealized system undulates in a quasiperiodic fashion in time;
this is the ``quasiperiodic dance'' referred to by Jorba and Simo.
Further support for this interpretation of Fig. 5 is provided by
the results shown in Fig. 6. In that figure, for the same model
system, finite time Lyapunov exponents are shown as a function of
initial condition for a set of air parcel trajectories that spans
the zonal jet. This figure shows that the region in the immediate
vicinity of the jet core is characterized by small Lyapunov
exponent estimates. This behavior is consistent with the
interpretation that there is a narrow band of surviving tori (on
which motion is nonchaotic) in this region.

\begin{figure}[t]
\centering
\includegraphics[width=14cm]{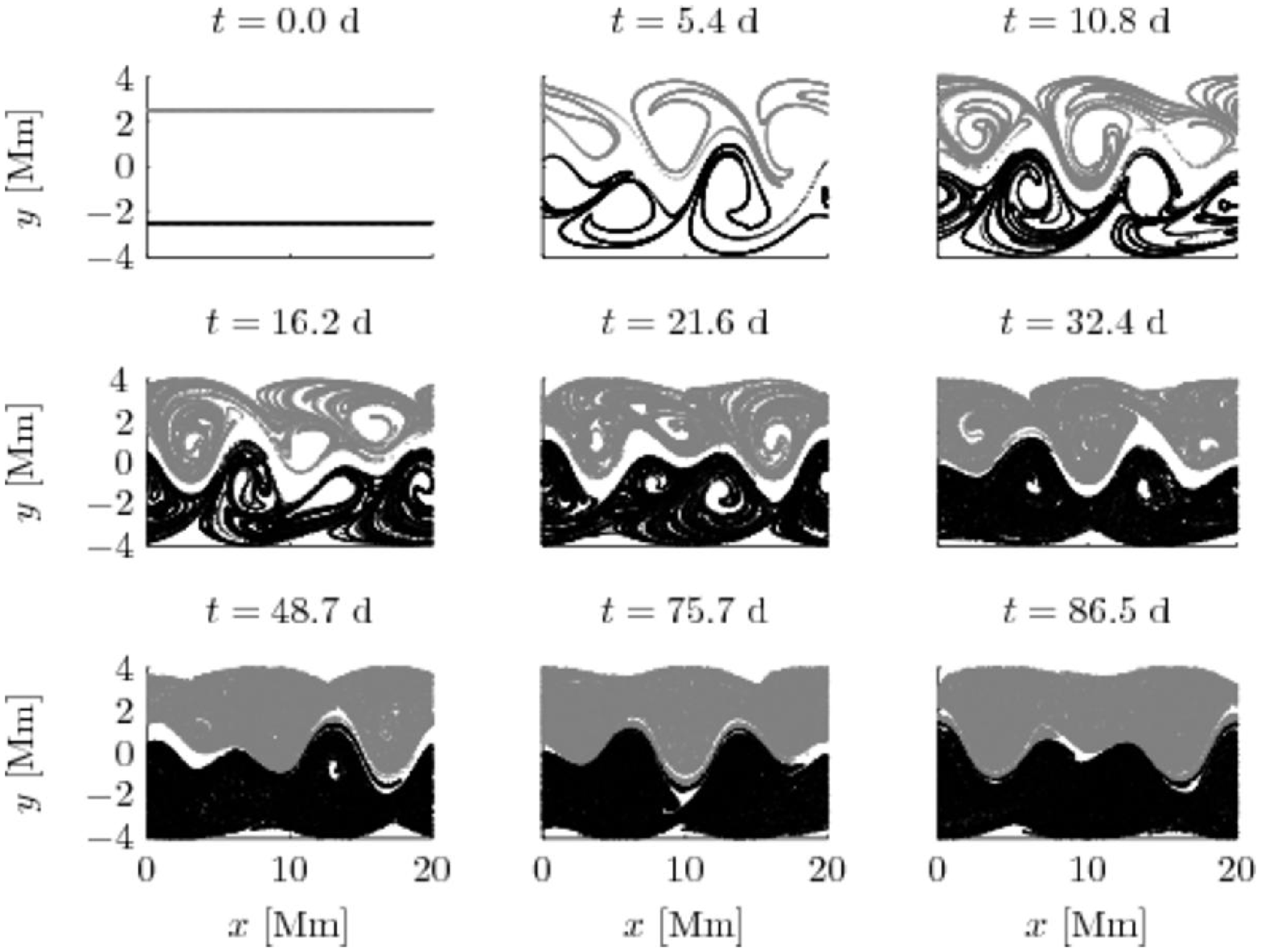}
\caption{Time evolution of two sets of 25000 points that at $t=0$
fall on zonal lines on opposite sides of the core of the zonal jet
in the system described by equations (\ref{eq:eq0}) and
(\ref{eq:eq17}) with $A_3 = 0.3$, $A_2 = 0.4$, $A_1 = 0.075$. Note
that, although trajectories are predominantly chaotic, there is no
transport across an undulating barrier in the vicinity of the jet
core.}
\end{figure}

\begin{figure}[t]
\centering
\includegraphics[width=14cm]{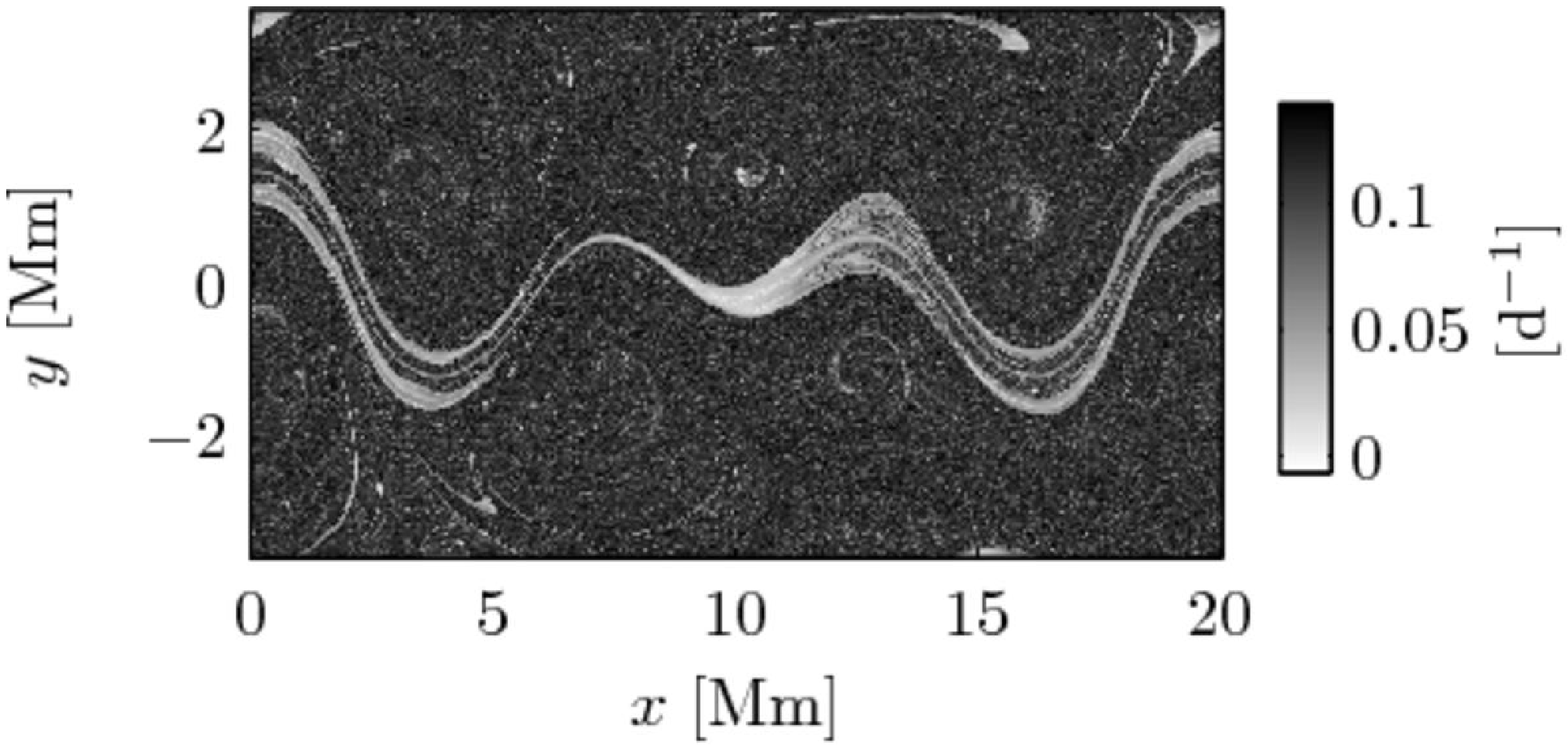}
\caption{Finite-time Lyapunov exponent estimates as a function of
initial position for the system described by equations
(\ref{eq:eq0}) and (\ref{eq:eq17}) with $A_3 = 0.3$, $A_2 = 0.4$,
$A_1 = 0.075$. The integration time for the estimates shown is
86.5 days.  Note that the region in the vicinity of the jet core
is characterized by small Lyapunov exponent estimates.}
\end{figure}

The qualitative features of Figs. 5 and 6 are consistent with
available observational evidence.  Consistent with Fig. 5 are
satellite-based measurements of ozone distributions in the austral
spring; see, e.g., Bowman and Mangus (1993)
\nocite{Bowman-Mangus-93} or the NASA/TOMS web site
(http://jwocky.gsfc.nasa.gov/eptoms/ep\_v8.html). These observations
consistently reveal a sharp boundary between low ozone concentration
air inside the stratospheric polar vortex and high ozone
concentration air outside the polar vortex. Aircraft-based
measurements \citep[see, e.g.,][]{Starr-Vedder-89} reveal an even
sharper boundary between these regions than is suggested by
satellite-based measurements; this is not surprising given that the
satellite measurements are integral measurements through the entire
atmosphere.  Our Fig. 6, which indicates that the perimeter of the
polar vortex is a narrow nonchaotic barrier that separates two
predominantly chaotic regions, is consistent with Fig. 8 in
\citet{Koh-Legras-02}, Fig. 2 in \citet{Pierce-Fairlie-93} and the
observation by \citet{Chen-94} that imbedded in the narrow barrier
between the inside and outside of the vortex is a potential
vorticity contour that grows at a locally minimal rate.

Figures 5 and 6 suggest that the most robust of the tori of the
original system are those in the vicinity of the core of the jet
where $\omega'(I) = 0$.  This observation is not surprising as it
is consistent with the discussion in the previous section relating
to resonance widths.  But the observation does serve to identify a
weakness in our argument, however, inasmuch as the Jorba--Simo
proof of the KAM theorem for quasiperiodic systems makes use of
the simplest (Kolmogorov) nondegeneracy condition $\omega'(I) \neq
0$. Thus the Jorba--Simo form of the KAM theorem does not address
the stability of tori near the jet core, i.e., those that are
apparently the most stable.  (One might argue that the theorem
holds for tori that are arbitrarily close to that for which
$\omega'(I) = 0$, but this is not entirely satisfactory in our
view given our focus on the jet core.)  What is needed to
rigorously complete our argument is a proof of the KAM theorem for
quasiperiodic systems (\ref{eq:eq16}) that makes use of a
Russmann-like nondegeneracy condition rather than the Kolmogorov
condition.  So far as we are aware, such a proof has not been
published to date. Our numerical simulations, including but not
limited to Figs. 5 and 6, strongly suggest that the theorem holds
for quasiperiodic systems (\ref{eq:eq16}) for which the background
$\omega'(I)$ has an isolated zero.

We turn our attention now to justifying the claim, made above
without proof, that for quasiperiodic systems (\ref{eq:eq16}) KAM
tori provide a barrier to transport.  Recall that for periodic
systems (\ref{eq:eq13}) this property was established by noting that
each KAM torus is a 2-dimensional surface in the 3-dimensional space
$(x,y,t \bmod 2\pi/\sigma)$ that divides the 3-d space into
nonintersecting ``inside'' and ``outside'' regions. An extension of
the same argument applies to the quasiperiodic problem.  To see
this, note first that the nonautonomous 1 degree-of-freedom system
described by equations (\ref{eq:eq0}) and (\ref{eq:eq16}) can be
written as an equivalent autonomous $N+1$ degree-of-freedom system,
\begin{equation}
\frac{dq_i}{d\tau} = \frac{\partial H}{\partial p_i}, \quad
\frac{dp_i}{d\tau} = -\frac{\partial H}{\partial q_i}, \quad i =
1,2,\ldots,N+1 \label{eq:eq18}
\end{equation}
where $q_i = \sigma_i t$, $p_i = -\psi/\sigma_i$, $i =
1,2,\ldots,N$, and $q_{N+1} = y$, $p_{N+1} = x$, with
\begin{equation}
H({\bf p},{\bf q}) = \psi(p_{N+1},q_{N+1};q_1,q_2,\ldots,q_N) +
\sum_{i=1}^N \sigma_i p_i. \label{eq:eq19}
\end{equation}
It is straightforward to verify that (\ref{eq:eq18}) and
(\ref{eq:eq19}) reduce to $dt/d\tau = 1$, equations (\ref{eq:eq0})
and $d\psi/dt = \partial \psi/\partial t$.  An important property of
the transformed system (\ref{eq:eq18}, \ref{eq:eq19}) is that each
trajectory is constrained by the presence of $N$ integrals
(sometimes called constants of the motion), i.e., $N$ functions
$f_i({\bf p},{\bf q})$, $i = 1,2,\ldots,N$ for which $df_i/d\tau =
0$.  These integrals are $H$ and $f_i = q_i/\sigma_i -
q_N/\sigma_N$, $i = 1,2,\ldots,N-1$.  If one additional independent
integral can be found, the system (\ref{eq:eq18}, \ref{eq:eq19}) can
be solved by quadratures and is said to be integrable.  (This should
come as no surprise because the original system (\ref{eq:eq0},
\ref{eq:eq16}) also lacks only one integral to render it
integrable.)  For our purposes, the principal significance of the
$N$ integrals is that, because of their presence, each trajectory in
the $2(N+1)$-dimensional phase space, lies on a surface of dimension
$2(N+1)-N = N+2$.  In a near-integrable system of this type in which
both KAM tori and chaotic trajectories are present, the tori have
dimension equal to the number of degrees of freedom, $N+1$. In the
$N+2$-dimensional space that are filled by chaotic trajectories, the
$N+1$-dimensional KAM tori serve as impenetrable transport barriers.
(The significance of these numbers is that the dimension of the KAM
tori is one less than the dimension of the space that the chaotic
trajectories fill.  Note, for example, that in $(x,y,z)$ the 1-d
circle $x^2+y^2=1, z=0$ divides the 2-d $z=0$ plane into
nonintersecting inside and outside regions, but the same 1-d circle
does not divide the 3-d $(x,y,z)$ volume into nonintersecting inside
and outside regions.)

The argument just given shows that in the system described by
(\ref{eq:eq18}) and (\ref{eq:eq19}) Arnold diffusion does not occur.
Loosely speaking, this is the process which allows chaotic
trajectories to bypass KAM invariant tori.  This process occurs in
near-integrable autonomous systems with $N \geq 3$ degrees of
freedom which, under perturbation, are constrained by only one
integral $H$. For such systems phase space has dimension $2N$,
chaotic trajectories lie on surfaces of dimension $2N-1$, while KAM
invariant tori have dimension $N$; for $N \geq 3$ these tori do not
serve as impenetrable barriers to the chaotic trajectories. The
cause of the absence of Arnold diffusion in the system described by
(\ref{eq:eq18}) and (\ref{eq:eq19}) is the integrals in addition to
$H$ that constrain the motion of all trajectories.

In this section we have argued that, with some minor
modifications, the conclusions of the previous section carry over
to a multiperiodic perturbation. Unlike the results of the
previous section, however, the multiperiodic argument lacks
mathematical rigor in that, to date, no proof of a KAM theorem for
quasiperiodically perturbed Russmann-nondegenerate Hamiltonians
has been published.  Numerical simulations provide strong evidence
that such a result holds.  With this in mind, we state with some
confidence that the qualitative features that were described in
the previous section -- the robust nature of nonchaotic
trajectories near the jet core that serve to isolate chaotic
trajectories on opposite sides of the jet -- are expected to be
seen whether there are 2 or 20 Rossby waves superimposed on the
background zonal jet.

\section{Summary and discussion}

In this paper we have argued, using several nontrivial extensions
of the basic KAM theorem, that, under commonly encountered
conditions, the zonal jet at the perimeter of the stratospheric
polar vortex provides a robust barrier to the meridional transport
of passive tracers.  In the model employed, the perturbation to
the background steady zonal jet was assumed to consist of a sum of
travelling Rossby waves.  The transport barrier is comprised of a
typically narrow band of nonchaotic trajectories, each lying on a
KAM invariant torus which is labelled by $I$, that survive in the
perturbed system. These tori tend to be the most resistant to
break-up under perturbation because they are close to the
unperturbed streamline near the jet core for which $\omega'(I)=0$
and because resonance widths $\Delta \omega$ are approximately
proportional to $|\omega'(I)|^{1/2}$. The required extensions to
the basic KAM theorem that we have made use of to arrive at this
conclusion are: 1) applicability of the theorem to multiperiodic
systems, including the absence of Arnold diffusion in such
systems; and 2) applicability of the theorem when the usual
(Kolmogorov) nondegeneracy condition $\omega'(I)=0$ is violated.
Our argument falls short of complete mathematical rigor because,
to our knowledge, published proofs of the KAM theorem for
quasiperiodically perturbed systems make use of the Kolmogorov
nondegeneracy condition, rather than the Russmann condition, as
required for our purposes.  (Note, however, that numerical
simulations strongly suggest that the theorem is satisfied for
Russmann-nondegenerate streamfunctions subject to quasiperiodic
perturbations, indicating that our argument is firmly, if not
rigorously, grounded.) Also, it should be emphasized that even
rigorous applicability of a form of the theorem to trajectories in
the vicinity of the jet core will not guarantee that for all
perturbations the corresponding tori will survive and provide a
barrier to meridional transport. KAM invariant tori may not
survive in the perturbed system for some combination of the
following reasons: (1) the phase speed of one of the more
energetic Rossby waves is close to the zonal velocity at the jet
core; (2) the perturbation excites a low-order resonance on one of
the tori in close proximity to that for which $\omega'(I)=0$; or
(3) the amplitude of the perturbation is too large.  In spite of
these caveats, our simulations suggest that, under conditions
similar to those found in the austral winter and spring, the
transport barrier near the core of the zonal jet at the perimeter
of the polar vortex is very robust.

Our KAM-theory-based explanation of the mechanism by which the
stratospheric polar vortex serves as a barrier to meridional
transport of passive tracers (such as ozone-depleted air within the
vortex) differs from the potential vorticity barrier argument
introduced originally by \citet[see also Juckes and McIntyre
1987\nocite{Juckes-McIntyre-87}]{McIntyre-89}. According to that
argument, on a given isentropic surface potential vorticity
(hereafter PV) is nearly constant following individual air parcels
and the PV distribution associated with the background zonal jet is
characterized by nearly uniform distributions on each side of the
jet separated by a region near the core of the jet within which the
PV gradient is strong. If the perturbation to the background PV is
sufficiently weak that the background meridional PV structure is
largely intact in the perturbed system, then PV conservation leads
to the expectation that the PV gradient maximum near the jet core
should serve to inhibit meridional transport, i.e., that the PV
gradient maximum serves as a ``PV-barrier.''

The PV-barrier argument has several weaknesses.  First, there is
no reason to expect that, in general, the vorticity distribution
associated with a zonal jet is characterized by a meridional
gradient with a prominent maximum near the jet core.  The
meridional gradient of the background relative vorticity is the
second derivative of $u_0(y)$.  It is easy to construct examples
of jet-like zonal velocity profiles whose second derivative does
not have a local maximum near the jet core -- a quadratic, for
example. Second, even when $u_0''(y)$ does have a peak near the
jet core, the PV-barrier argument holds only for a very weak
perturbation. Let $U$ and $L$ denote characteristic velocity and
length scales with the subscripts $0$ and $1$ used to denote
background and perturbation, respectively. The ratio of the
magnitude of the relative vorticity of the perturbation to that in
the background is $|\zeta_1/\zeta_0| = O((L_0/L_1)(U_1/U_0))$.
Under typical ozone-trapping conditions in the stratosphere this
ratio is not small owing to the fact that $L_0/L_1$ is greater
than unity. Under typical trapping conditions $\zeta_1/\zeta_0$ is
order unity.  Under such conditions the relative vorticity
structure of the background flow is not expected to strongly
constrain the perturbed system. Third, even when the background
zonal jet is characterized by a relative vorticity distribution
with a maximum meridional gradient near the jet core and the
perturbation is very weak, this argument suggests that the
transport barrier should be broad, diffuse and stationary
(centered at the latitude of the maximum background vorticity
gradient), as opposed to being a narrow, wobbly nearly impermeable
barrier. Observational evidence supports the latter view.  And
fourth, the PV-barrier argument provides no insight into why the
transport barrier breaks down (as is readily confirmed in
simulations) when one of the dominant Rossby wave phase speeds is
comparable to the jet core velocity, or when a low order resonance
is excited by one of the dominant Rossby waves. In contrast, our
KAM-theory-based argument: (1) is robust inasmuch as it requires
only that $u_0(y)$ has a local maximum, but assumes nothing about
$u_0''(y)$; (2) holds when $|\zeta_1/\zeta_0|$ is $O(1)$ (the KAM
theorem assumes that the perturbation is small but numerical
simulations reveal that provided no low-order resonances are
excited, KAM tori survive even when $(U_1/U_0)$ is $O(1)$); (3)
naturally explains the occurrence of a narrow impermeable barrier
that wobbles in the vicinity of the jet core, and the observation
that within this narrow region neighboring trajectories diverge
from one another only very slowly; and (4) naturally accounts for
the breakdown of the barrier when one of the dominant Rossby wave
phase speeds is comparable to the jet core velocity, or when a low
order resonance is excited by one of the dominant Rossby waves.

The foregoing arguments should not be interpreted as an assertion
that the PV-barrier argument is incorrect.  Rather, we are arguing
that a PV-barrier is not a necessary condition for trapping of air
inside the polar vortex, and that a barrier of this type at the
perimeter of the polar vortex is probably also not typical.  The
latter point is supported by the analysis, based on analyzed
winds, of \citet{Paparella-etal-97}. Figure 2 in that paper shows
that the vortex edge is often not associated with a strong
meridional PV gradient, but that the vortex edge does appear to be
reliably identified as a maximum of kinetic energy. This behavior
is in good qualitative agreement with our arguments (recall our
Fig. 3) in that in the background environment the trajectory for
which $\omega'(I) = 0$ is close to that for which the kinetic
energy is maximum.

The transport barrier at the perimeter of the stratospheric polar
vortex that we have identified as being due to a thin band of KAM
invariant tori can be described as a Lagrangian coherent
structure. The subject of Lagrangian coherent structures has been
extensively studied in recent years \citep[see,
e.g.,][]{Malhotra-Wiggins-98,Haller-00,Haller-Yuan-00,Haller-01,Haller-02,Shadden-etal-05}.
In most applications the Lagrangian coherent structures of
interest are the stable and/or unstable manifolds of perturbed
hyperbolic points.  Unlike the KAM tori in our study, which
constitute global barriers for transport, such invariant manifolds
are barriers for transport only in a local sense and for
sufficiently short time.  Also, while KAM tori are associated with
regular motion, the stable and unstable manifolds are generically
associated with chaotic motion in the vicinity of their points of
intersections (homoclinic points).

A natural and important extension of the mostly theoretical work
reported here is to use analyzed winds \citep[following, e.g.,][or
Koh and Legras 2002\nocite{Koh-Legras-02}]{Bowman-93,Bowman-96} to
more thoroughly test the predictions made here versus those of the
PV-barrier paradigm. An empirical study of this type must employ
spherical coordinates, i.e., $\psi = \psi(\lambda, \varphi, t)$ on
a selected isentropic surface where $\lambda$ and $\varphi$ are
longitude and latitude, respectively. Questions that could be
addressed with such a model include the following. Is our
hypothesized rest frame decomposition of $\psi$, $\psi(\lambda,
\varphi, t) = \psi_0(\varphi) + \psi_1(\lambda, \varphi, t)$ where
$\psi_1$ is a superposition of zonally propagating Rossby waves, a
good approximation? Are vorticity distributions consistent with
the PV-barrier paradigm? Is the transport barrier a thin wobbly
region on which Lagrangian motion is nonchaotic, as we predict? Is
the transport barrier associated with a maximum PV gradient, a
maximum of kinetic energy, or something else? Is the breakup of
the transport barrier on a given isentropic surface caused by
either the phase speed of one of the dominant Rossby waves being
comparable to the wind speed at the jet core or the excitation of
a low-order resonance, as we have suggested?

\acknowledgement This work was supported by the U.S. National
Science Foundation, grant CMG0417425.

\bibliographystyle{ametsoc}

\newpage
\section*{Figure Captions}

\noindent \textsc{Fig.} 1.  For the streamfunction $\psi_0(y) = -U_0
L \tanh(y/L)$: (a) selected level surfaces of $\psi_0$; (b)
$u_0(y)$; (c) $\omega(I)$; (d) $T(I)$; (e) $\omega'(I)$.  In this
and subsequent figures, d denotes days and distance is measured in Mm;
1 Mm = 1000 km. \\

\noindent \textsc{Fig.} 2.  Bifurcation diagram in the $(A,c/U_0)$
parameter space corresponding to the streamfunction $\psi(x,y) =
cy - U_0 L \tanh(y/L) + A U_0 L {\rm sech}^2(y/L)\cos(kx)$.  There
are three topologically distinct regions and two critical curves
separating these regions. Selected level surfaces of $\psi(x,y)$
in each of the three regions and on the two critical curves are
shown. In the level surface plots zonal wavenumber three is
assumed so $k = k_3$, and $0 \leq k_3x \leq 6\pi$.\\

\noindent \textsc{Fig.} 3.  For the streamfunction $\psi(x,y) = c_3y
- U_0 L \tanh(y/L) + A_3 U_0 L {\rm sech}^2(y/L)\cos(k_3x)$ with
$A_3 = 0.3$, $c_3/U_0 = 0.461$: (a) selected level surfaces of
$\psi(x,y)$; (b) $|{\bf u}|(y)$ at $k_3x = \pi/2$; (c) $\omega(I)$;
(d) $T(I)$; and (e) $\omega'(I)$. In (b), (c), (d) and (e) only
values of $y$ and $I$ corresponding to the shaded region near the
jet core in (a) are shown.\\

\noindent \textsc{Fig.} 4. Poincare sections corresponding to the
system described by equations (\ref{eq:eq0}) and (\ref{eq:eq12})
with $A_3 = 0.3$ for three values of $A_2$: $0$ (upper plot),
$0.1$ (middle plot), and $0.7$ (lower plot).  Note the robustness
of the tori in the vicinity of the jet core.\\

\noindent \textsc{Fig.} 5. Time evolution of two sets of 25000
points that at $t=0$ fall on zonal lines on opposite sides of the
core of the zonal jet in the system described by equations
(\ref{eq:eq0}) and (\ref{eq:eq17}) with $A_3 = 0.3$, $A_2 = 0.4$,
$A_1 = 0.075$.  Note that, although trajectories are predominantly
chaotic, there is no transport across an undulating barrier in the
vicinity of the jet core.\\

\noindent \textsc{Fig.} 6. Finite-time Lyapunov exponent estimates
as a function of initial position for the system described by
equations (\ref{eq:eq0}) and (\ref{eq:eq17}) with $A_3 = 0.3$, $A_2
= 0.4$, $A_1 = 0.075$. The integration time for the estimates shown
is 86.5 days.  Note that the region in the vicinity of the jet core
is characterized by small Lyapunov exponent
estimates.\\

\end{article}
\end{document}